\documentclass[fleqn,twoside]{article}
\usepackage{espcrc2}

\usepackage{graphicx}
\usepackage{epsfig}

\newcommand{\Dlr}{\stackrel{\leftrightarrow}{D}}
\newcommand{\Dl}{\stackrel{\leftarrow}{D}}
\newcommand{\Dr}{\stackrel{\rightarrow}{D}}
\newcommand{\HG}{H(4)}
\newcommand{\cO}{\mathcal {O}}
\newcommand{\gr}{g_{\mathrm R}}
\newcommand{\quarter}{\mbox{\small $\frac{1}{4}$}}
\newcommand{\MS}{{\overline{\rm MS}}}

\title{
\thispagestyle{empty}
\vspace{-17mm}
\rightline
{\small DESY 04-165, Edinburgh 2004/15, LU-ITP 2004/020, LTH 632}
\rightline{\small  September  2004}
\vspace{5mm}
One-loop renormalisation for the second moment
of GPDs with Wilson fermions
}

\author{M.~G\"ockeler\address{Institut f\"ur Theoretische Physik, Universit\"at
Leipzig, D-04109 Leipzig, Germany}$^,$\address{Institut f\"ur Theoretische
Physik, Universit\"at Regensburg, D-93040 Regensburg, Germany},
R.~Horsley\address{School of Physics, University of Edinburgh,
                   Edinburgh EH9 3JZ, UK},
H.~Perlt$^{\rm b,a}$,
P.~E.~L.~Rakow\address{Department of Mathematical Sciences,
            University of Liverpool, Liverpool L69 3BX, UK},
A.~Sch\"afer$^{\rm b}$,
G.~Schierholz\address{John von Neumann-Institut f\"ur Computing NIC,
DESY Zeuthen, D-15738 Zeuthen, Germany}%
$^,$\address{Deutsches Elektronen-Synchrotron DESY, D-22603 Hamburg, Germany},
A.~Schiller$^{\rm a,}$\thanks{Talk presented by A.~Schiller
at Lattice 2004, Fermilab, USA, June 21-26,2004.}
}

\begin{document}

\begin{abstract}
We calculate the non-forward quark matrix elements for operators with two
covariant derivatives in one-loop lattice perturbation theory using Wilson
fermions. These matrix elements are needed in the renormalisation
of the second moment of generalised parton distributions measured in
lattice QCD. For some commonly used representations of the hypercubic group
we determine the sets of all mixing operators and find the matrices of mixing
and renormalisation factors.
\vspace{-5mm}
\end{abstract}

\maketitle

In recent years generalised parton distributions (GPDs) have been intensively
studied both in experiments and theoretically~\cite{Diehl} to extend our knowledge
of hadron structure. However, the direct experimental access to GPDs beyond the
limiting cases of distribution functions and simple form factors is limited.
Therefore, it is indispensable  to obtain complementary information from lattice
QCD calculations.

On the lattice we can compute matrix elements of local composite operators, and
moments of GPDs can be related to such matrix elements taken between states of
different nucleon momenta and spins. First results for moments of GPDs have been
published recently~\cite{QCDSFHagler}, see also~\cite{lat04}, and soon results from
improved calculations should become available.

In order to relate lattice measurements to continuum quantities we have to
investigate the renormalisation and mixing of the operators involved.
When non-forward matrix elements are studied, new features arise, which make a
reconsideration of the renormalisation problem necessary. In particular,
the mixing with ``external ordinary derivatives'', i.e.\ with operators
of the form $\partial_\mu \partial_\nu \cdots (\bar{\psi} \cdots \psi)$,
needs to be taken into account.
On the lattice the mixing patterns are usually more complicated than
in the continuum, because covariance under the hypercubic group $\HG$
imposes less stringent restrictions than $O(4)$ covariance. The necessity
to consider also external ordinary derivatives enlarges the set of contributing
operators even further.
These complications do not yet arise for operators with zero or one covariant derivative.
Hence for these moments the renormalisation factors can be taken
over from the forward case.

Here we investigate the renormalisation problem for operators with two
covariant derivatives sandwiched between off-shell quark states at different
momenta within the framework of one-loop lattice perturbation theory and report
on first results~\cite{preparation}. We use the Wilson plaquette gauge action
with Wilson fermions and perform the calculations in Feynman gauge.

Let us first discuss renormalisation and mixing in general.
Denote by $\Gamma_j(p',p,\mu,\gr,\epsilon)$ (with $j=1,2,\ldots,N$) the
dimensionally regularised amputated vertex functions of $N$ mixing operators
$\cO_j$, $p$ and $p'$ are the quark momenta.
The corresponding Born terms are denoted by $\Gamma_j^{\mathrm {B}}(p',p)$.
The renormalised coupling constant $\gr$ is related to the bare coupling  $g$ by
$\gr^2 = \mu ^{-2\epsilon} g^2 \left( 1 + O(g^2) \right)$,
where $\mu$ is the renormalisation scale.
In the $\MS$ scheme the corresponding renormalised vertex functions are given by
\vspace{-3mm}
\begin{equation}
  \label{msvf}
  \begin{array}{l} \displaystyle
    \Gamma_j^R (p',p,\mu,\gr) = \Gamma_j^{\mathrm {B}}(p',p)
    + \gr^2 \Big[ \sum_{k=1}^N (- \gamma_{jk}^V)
    \\ \displaystyle
     {} \times
    \ln \frac{(p'+p)^2}{4 \mu^2}
    \Gamma_k^{\mathrm {B}}(p',p)
    + f_j (p',p) \Big]
    \,,
  \end{array}
\end{equation}
where $f_j$ is the finite one-loop contribution.
All $O(\gr^4)$ contributions are neglected.

In the absence of mixing with lower-dimensional operators the
vertex functions regularised on a lattice with lattice spacing $a$ are
\vspace{-3mm}
\begin{equation}
  \label{latvf}
  \begin{array}{l} \displaystyle
     \Gamma_j^L (p',p,a,\gr) = \Gamma_j^{\mathrm {B}}(p',p) + \gr^2 
     \Big[ \sum_{k=1}^N (- \gamma_{jk}^V)
     \\ \displaystyle
     {} \times
     \ln \left( \quarter a^2 (p'+p)^2 \right) \Gamma_k^{\mathrm {B}}(p',p)
     + f_j^L (p',p) \Big] \,.
  \end{array}
\end{equation}
The relation between $\Gamma_k^L $ and the $\MS$
renormalised vertex functions $\Gamma_j^R$ is given by
\vspace{-2mm}
\begin{equation}
  \label{zeta}
  \Gamma_j^R =
  \sum_{k=1}^N \left( \delta_{jk} + \gr^2 \zeta_{jk}
  \right)
  \Gamma_k^L  \,,
\end{equation}
\vspace{-2mm}
where the matrix  $\zeta$ is found  to be
\begin{equation}
  \label{zeta2}
  \vspace{-2mm}
  \zeta_{jk} = \gamma_{jk}^V \ln \left( a^2 \mu^2 \right) - c_{jk}^V
\end{equation}
\vspace{-1mm}
with the constants $c_{jk}^V$ determined from
\vspace{-2mm}
\begin{equation}
  \label{const}
  f_j^L (p',p) - f_j (p',p) =
  \sum_{k=1}^N c_{jk}^V \Gamma_k^{\mathrm {B}}(p',p) \,.
\end{equation}
\vspace{-2mm}
Mixing with lower-dimensional operators leads to the appearance of
additional terms on the r.h.s.\ of Eq.~(\ref{latvf}), for details
see~\cite{preparation}.

The operators potentially contributing to the mixing have to
transform identically according to a given irreducible
representation of  $O(4)$ or $\HG$ and should have the same charge
conjugation parity.

In terms of the quark wave function renormalisation constant $Z_\psi$ and
the matrix $Z_{jk}$ of mixing and renormalisation coefficients,
the connection between the bare lattice vertex functions and the $\MS$
renormalised vertex functions can be written as
\vspace{-2mm}
\begin{equation}
  \Gamma_j^R = Z_\psi^{-1}
  \sum_{k=1}^N Z_{jk} \Gamma_k^L \,.
\end{equation}
Using the known $Z_\psi$,
\vspace{-2mm}
\begin{equation}
  Z_\psi = 1 - \frac{\gr^2 \,C_F}{16 \pi^2}
  \left[ \ln \left( a^2 \mu^2 \right) + 1 + \sigma_L \right]
\end{equation}
with  $\sigma_L=11.8524$ for Wilson fermions, we find
\begin{equation} \label{zjk}
  Z_{jk} = \delta_{jk} -
  \frac{\gr^2 \,C_F}{16\pi^2}
  \left[\gamma_{jk}\,\log(a^2\mu^2) + c_{jk} \right]
\end{equation}
with
$ \gamma_{jk}=\delta_{jk}- ({16 \pi^2}/{C_F}) \gamma_{jk}^V$
and
$c_{jk}=\delta_{jk} (1+\sigma_L )+({16 \pi^2}/{C_F}) c_{jk}^V$.

\begin{table*}[!htb]
$$
  c_{jk} =\left( \begin{array}{rrrrrr}
  -12.1274 & 1.4913       & 0.3685      & -0.4160     & 0.0156   & 0.1498\\
     0     &  20.6178     & 0           & 0           & 0        & 0\\
   3.3060  & -8.01456     & -14.8516    &  4.3023     & -0.9285  & 0.7380 \\
     0     & 0            & 0           &  20.6178    & 0        & 0 \\
  0        & 3.2644       & 0           & 0           & 0.3501   & 0.0149 \\
  0        & 3.2644       & 0           & 0           & 0.0050   & 0.3600
  \end{array}
  \right)
$$
\vspace{-4mm}
\caption{The finite mixing part for the operators $\cO_1, \dots, \cO_6$
in (\ref{O1}), (\ref{O2345}).}
\label{tab1}
\end{table*}

The one-loop computation has been performed  symbolically, adopting and
significantly extending a {\it Mathematica} program package developed originally
for the case of moments of structure functions using Wilson ~\cite{Gockeler:1996hg},
clover~\cite{Capitani:2001xi} and overlap fermions~\cite{Horsley:2004mx}.
Using that approach we have full analytic control over pole cancellation.
The Lorentz index structure of the matrix elements is left completely free,
so that we are able to construct all representations of $\HG$ for the second
moments in the non-forward case.

We have to deal with operators of the following general types
($\Dlr=\Dr-\Dl$):
\begin{equation}
  \label{xyz}
  \begin{array}{l} \displaystyle
    \cO_{\mu\nu\omega}^{DD}= (-1/4)\,
    \bar\psi \gamma_\mu \Dlr_\nu \Dlr_\omega\psi\,,
    \vspace{1mm}
    \\ \displaystyle
    \cO_{\mu\nu\omega}^{\partial D}=  (-1/4)\,
    \partial_\nu \left( \bar\psi \gamma_\mu \Dlr_\omega\psi \right) \,,
    \vspace{1mm}
    \\ \displaystyle
    \cO_{\mu\nu\omega}^{\partial \partial}= (-1/4)\,
    \partial_\nu \partial_\omega
    \left(\bar\psi \gamma_\mu \psi \right)\,,
    \vspace{1mm}
    \\ \displaystyle
    \cO_{\mu\nu\omega}^\partial=(-i/2)\,
    \partial_\omega
    \left( \bar\psi  [\gamma_\mu,\gamma_\nu ]  \psi \right)
  \end{array}
\end{equation}
and similarly for operators with $\gamma_\mu \gamma_5$.
In lattice momentum space we realise the operators with non-zero momentum transfer
$q$ (showing as an example an operator with one covariant derivative) through
\vspace{-2mm}
\begin{equation}
  \label{DII}
  \begin{array}{l} \displaystyle
     \hspace{-1mm}
     \left (\bar{\psi}\Dlr_\mu \psi\right)\!\!(q)
     =
     \frac{1}{a} \sum_x
     \,{\rm e}^{i q \cdot(x+a\hat{\mu}/2)} \times
     \\ \displaystyle
     \hspace{-2mm}
     \left[\bar{\psi}(x)U_{x,\mu} \psi(x+a\hat{\mu})-
     \bar{\psi}(x+a\hat{\mu})U^\dagger_{x,\mu} \psi(x)\right]
     \,.
  \end{array}
\end{equation}
Eq.~(\ref{DII}) basically defines the Feynman rules in lattice
perturbation theory.

Consider now the following operator:
\vspace{-2mm}
\begin{equation}
  \cO_1=\cO^{DD}_{\{114\}}-\frac{1}{2}
  \left(\cO^{DD}_{\{224\}}+\cO^{DD}_{\{334\}}\right)
  \label{O1}
\end{equation}
with charge conjugation parity $C=-1$  and corresponding  to the representation
$\tau^{(8)}_1$ of $\HG$. We use the notations~\cite{group}:
\begin{equation}
  \begin{array}{l}\displaystyle
     \cO_{ \{ \nu_1\nu_2\nu_3 \} }= (1/6)\, (
     \cO_{\nu_1\nu_2\nu_3}+\cO_{\nu_1\nu_3\nu_2}
     \\ \displaystyle
     \ \  + \cO_{\nu_2\nu_1\nu_3} +
     \cO_{\nu_2\nu_3\nu_1}+\cO_{\nu_3\nu_1\nu_2}+\cO_{\nu_3\nu_2\nu_1}  )
     \,,
     \\ \displaystyle
     \cO_{\|\nu_1\nu_2\nu_3\| } = \cO_{\nu_1\nu_2\nu_3}-\cO_{\nu_1\nu_3\nu_2}
     + \cO_{\nu_3\nu_1\nu_2}
     \\ \displaystyle
     \ \   -\cO_{\nu_3\nu_2\nu_1}-2\,\cO_{\nu_2\nu_3\nu_1}
     +2\,\cO_{\nu_2\nu_1\nu_3}
     \,,
     \\ \displaystyle
     \cO_{\langle\langle\nu_1\nu_2\nu_3\rangle\rangle } =
     \cO_{\nu_1\nu_2\nu_3}+\cO_{\nu_1\nu_3\nu_2}
     \\ \displaystyle
     \ \  -\cO_{\nu_3\nu_1\nu_2}-\cO_{\nu_3\nu_2\nu_1}
     \,.
  \end{array}
\end{equation}
The following operators (transforming identically) mix in one-loop order with
$\cO_1$:
\vspace{-1mm}
\begin{equation}
  \label{O2345}
   \begin{array}{l} \displaystyle
      \cO_2=\cO^{\partial\partial}_{\{114\}}-  \left(
      \cO^{\partial\partial}_{\{224\}}
      +\cO^{\partial\partial}_{\{334\}}\right)/2
      \,,
      \\ \displaystyle
      \cO_3=\cO^{DD}_{\langle\langle 114\rangle\rangle}-
      \left( \cO^{DD}_{\langle\langle224\rangle\rangle}+
      \cO^{DD}_{\langle\langle334\rangle\rangle}\right)/2
      \,,
      \\ \displaystyle
      \cO_4=\cO^{\partial\partial}_{\langle\langle 114 \rangle\rangle}-
      \left(  \cO^{\partial\partial}_{\langle\langle 224 \rangle\rangle}+
      \cO^{\partial\partial}_{\langle\langle 334 \rangle\rangle}\right)/2
      \,,
      \\ \displaystyle
      \cO_5=\cO^{5,\partial D}_{||213||}
      \,,
      \quad
      \cO_6=\cO^{5,\partial D}_{\langle\langle213\rangle\rangle}
  \end{array}
\end{equation}
together with the lower-dimensional operator
\begin{equation}
  \label{O8}
  \cO_8= \cO^\partial_{411}-
  \left(\cO^\partial_{422}+\cO^\partial_{433} \right)/2
  \,.
\end{equation}
An additional operator with the same symmetry properties does not contribute
in the one-loop approximation. The anomalous dimension matrix is obtained as
\begin{equation}
   \gamma_{jk} =\left( \begin{array}{rrrrrr}
   \frac{25}{6}&-\frac{5}{6}& 0          & 0          & 0          & 0\\
      0        & 0          & 0          & 0          & 0          & 0\\
      0        & 0          & \frac{7}{6}&-\frac{5}{6}&1           &-\frac{3}{2}\\
      0        & 0          & 0          & 0          & 0          & 0 \\
      0        & 0          & 0          & 0          & 2          & -2 \\
      0        & 0          & 0          & 0          & -\frac{2}{3}& \frac{2}{3}
  \end{array}
  \right)
  \label{andim2}
\end{equation}
and the finite mixing matrix part is given in Table~\ref{tab1}.
There is an additional mixing of $\cO_1$ with the lower-dimensional operator
$\cO_8$ \ (in Born approximation):
\begin{equation}
  \cO_1\bigg|_{\frac{1}{a}-{\rm part}} =
  \frac{\gr^2\, C_F}{16\pi^2}(-0.5177)\,\frac{1}{a}\, \cO_8^{\rm B}\,.
\end{equation}
This mixing leads to a contribution which potentially diverges like
$1/a$ in the continuum limit.
The operator $\cO_8$ has to be subtracted non-perturbatively from the
matrix element of the operator $\cO_1$  which might be a difficult task
in simulations.

In summary, we have presented first results for the one-loop quark
matrix elements of operators needed for the second moments of GPDs.
This allows us to determine the mixing matrix of $Z$-factors in the $\MS$-scheme
for operators commonly used in simulations.
In the case of the representation $\tau_1^{(8)}$, a sizeable mixing
with operators of the same dimension appears. Moreover, mixing with a
lower-dimensional operator has to be taken into account.
Additional results, in particular for {\sl transversity} operators, will be
presented in~\cite{preparation}.

This work has been supported by DFG under contract FOR 465
(Forschergruppe Gitter-Hadronen-Ph\"{a}nomeno\-logie).

\end{document}